\documentstyle[amsfonts,12pt,epsfig]{article}

\input{tcilatex}

\begin{document}

\onecolumn

\begin{titlepage}
\begin{center}
{\LARGE \bf Chaos in a Relativistic 3-body Self-Gravitating System}
\\ \vspace{2cm}
F. Burnell\footnotemark\footnotetext{email: 
fburnell@physics.ubc.ca}
R.B. Mann\footnotemark\footnotetext{email: 
mann@avatar.uwaterloo.ca},
\\
\vspace{0.5cm} 
Dept. of Physics,
University of Waterloo
Waterloo, ONT N2L 3G1, Canada\\
\vspace{1cm}
T. Ohta \footnotemark\footnotetext{email:
t-oo1@ipc.miyakyo-u.ac.jp}\\
\vspace{0.5cm} 
Department of Physics, Miyagi University of Education,
Aoba-Aramaki, Sendai 980, Japan\\
\vspace{2cm}
\vspace{2cm}
\today\\
\end{center}

\begin{abstract}
We consider the 3-body problem in relativistic lineal (ie 1+1 dimensional) gravity 
and obtain an 
exact expression for its Hamiltonian and equations of motion. 
While general-relativistic effects yield more tightly-bound orbits of higher 
frequency compared to their non-relativistic counterparts, as energy increases 
we find in the equal-mass case no evidence for either global chaos or a 
breakdown from regular to chaotic motion, despite the high degree of 
non-linearity in the system. We find numerical evidence for mild chaos and 
a countably 
infinite class of non-chaotic orbits, yielding a fractal structure in the 
outer regions of the Poincare plot.
\end{abstract}
\end{titlepage}\onecolumn

Self-gravitating $N$-body systems have long \ been of key importance in
studying both stellar dynamics (small $N\geq 2$) and galactic evolution
(large $N$). Solutions to their problem of motion, complicated for $N>2$
even in nonrelativistic mechanics due to singularities and evaporation, are
essentially intractable for $N>1$ in general relativity because of
gravitational radiation. These difficulties are avoided in one-dimensional
(or lineal) models of such systems, which describe a system of $N$ parallel
mass sheets. Studied by astronomers and physicists for several decades \cite%
{Rybick}, they merit special attention \cite{yawn,rogs}, since they not only
admit a significantly greater level of computational and analytic analysis
but can also be mapped to systems subject to 
experimental test \cite{optbill}.

We report here on the results of an investigation of the $3$-body problem
for a relativistic self-gravitating lineal system. Three is the smallest
number of particles for which chaotic motions are possible and so such
systems are of particular interest. Non-relativistically this system is
equivalent to that of a single particle moving in a hexagonal funnel \cite%
{Butka} or to that of a uniformly accelerated particle undergoing elastic
collisions with a wedge \cite{Lmiller}. While its dynamics are
well-documented, ours is the first non-perturbative relativistic treatment
of this problem. From the canonical formulation of lineal gravity minimally
coupled to $3$ particles, we obtain an exact expression for the Hamiltonian
(valid to all orders in the gravitational coupling $\kappa =8\pi G/c^{4}$),
transcendentally expressed as a function of the four independent degrees of
freedom. From this we are able to solve for the motion of the system under
arbitrary initial conditions and study quasi-periodic motion and chaos in a
general-relativistic self-gravitating system.

For our lineal self-gravitating system we minimally couple $N$ point masses
to gravity in 2 spacetime dimensions%
\begin{equation}
I=\int d^{2}x\left[ \frac{\sqrt{-g}}{2\kappa }\left\{ \Psi R+\frac{1}{2}%
g^{\mu \nu }\nabla _{\mu }\Psi \nabla _{\nu }\Psi +\Lambda \right\}
-\sum_{a=1}^{N}m_{a}\int d\tau _{a}\left\{ -g_{\mu \nu }(x)\frac{dz_{a}^{\mu
}}{d\tau _{a}}\frac{dz_{a}^{\nu }}{d\tau _{a}}\right\} ^{1/2}\delta
^{2}(x-z_{a}(\tau _{a}))\right]  \label{eq1}
\end{equation}%
where $g_{\mu \nu }$ and $g$ are the metric and its determinant, $R$ is the
Ricci scalar, $\tau _{a}$ is the proper time of $a$-th particle, and we
incorporate a scalar (dilaton) field $\Psi $ since the Einstein action is a
topological invariant in $2$ spacetime dimensions. \ \ The action (\ref{eq1}%
) describes a generally covariant self-gravitating system (without
collisional terms, so that the bodies pass through each other) that is a
generalization of \ Jackiw-Teitelboim lineal gravity \cite{JT}, in which the
scalar curvature is equated to a (cosmological) constant $\Lambda $; its
equations of motion are \noindent 
\begin{equation}
R-\Lambda =\kappa T_{\;\;\mu }^{P\mu }\quad \frac{d}{d\tau _{a}}\left\{ 
\frac{dz_{a}^{\nu }}{d\tau _{a}}\right\} +\Gamma _{\alpha \beta }^{\nu
}(z_{a})\frac{dz_{a}^{\alpha }}{d\tau _{a}}\frac{dz_{a}^{\beta }}{d\tau _{a}}%
=0  \label{RTgeo}
\end{equation}%
\begin{equation}
\frac{1}{2}\nabla _{\mu }\Psi \nabla _{\nu }\Psi -g_{\mu \nu }\left( \frac{1%
}{4}\nabla ^{\lambda }\Psi \nabla _{\lambda }\Psi -\nabla ^{2}\Psi \right)
-\nabla _{\mu }\nabla _{\nu }\Psi =\kappa T_{\mu \nu }^{P}+\frac{\Lambda }{2}%
g_{\mu \nu }  \label{e4}
\end{equation}%
where the stress-energy due to the point masses is 
\begin{equation}
T_{\mu \nu }^{P}=\sum_{a=1}^{N}m_{a}\int d\tau _{a}\frac{1}{\sqrt{-g}}g_{\mu
\sigma }g_{\nu \rho }\frac{dz_{a}^{\sigma }}{d\tau _{a}}\frac{dz_{a}^{\rho }%
}{d\tau _{a}}\delta ^{2}(x-z_{a}(\tau _{a}))  \nonumber
\end{equation}%
and is conserved. Eq. (\ref{RTgeo}) is a closed system of $N+1$ equations
for which one can solve for the single metric degree of freedom and the $N$
degrees of freedom of the point masses. The evolution of the dilaton field
is governed by the evolution of the point-masses via (\ref{e4}). The
left-hand side of (\ref{e4}) is divergenceless (consistent with the
conservation of $T_{\mu \nu }^{P}$), yielding only one independent equation
to determine the single degree of freedom of the dilaton.

For general $N$ the canonical formulation of (\ref{eq1}) has been studied
previously \cite{ohtarobb} and a number of exact solutions to the $2$-body
problem have been obtained \cite{exact2body}. \ Writing the metric as $%
ds^{2}=-\left( N_{0}dt\right) ^{2}+\gamma \left( dx+\frac{N_{1}}{\gamma }%
dt\right) ^{2}$(with $\int dx\sqrt{\gamma }$\ describing proper spatial
distance at fixed $t$), the action can be canonically reduced to $I=\int
d^{2}x\left\{ \sum_{a=1}^{N}p_{a}\dot{z}_{a}\delta (x-z_{a})-{\cal H}%
\right\} $, where the reduced Hamiltonian is $H=\int dx{\cal H}=-\frac{1}{%
\kappa }\int dx\Psi ^{\prime \prime }$ , with the overdot and prime denoting 
$\frac{\partial }{\partial t}$ and $\frac{\partial }{\partial x}$\
respectively. The field $\Psi =\Psi (x,z_{a},p_{a})$ is understood to be
determined from the constraint equations, which become%
\begin{eqnarray}
\Psi ^{\prime \prime }-\frac{(\Psi ^{\prime })^{2}}{4}+\kappa ^{2}\left(
\chi ^{\prime }\right) ^{2}-\frac{\Lambda }{2}+\kappa \sum_{a}\sqrt{%
p_{a}^{2}+m_{a}^{2}}\delta (x-z_{a}) &=&0  \label{psicon} \\
2\chi ^{\prime \prime }+\sum_{a}p_{a}\delta (x-z_{a}) &=&0\;.  \label{picon}
\end{eqnarray}%
where $\pi =\chi ^{\prime }$ and $p_{a}$ are the respective momenta
conjugate to $\gamma $ and $z_{a}$. We have chosen the momentum conjugate to 
$\Psi $\ to vanish and $\gamma =1$, \ fixing the frame of the physical
space-time coordinates in a manner analogous to the $(3+1)$-dimensional case %
\cite{adm}.

Remarkably eqs. (\ref{psicon},\ref{picon}) can be exactly solved when $N=3$.
The solution for $\Psi $ is a sum of exponentials, and for $\chi $ a linear
function of $x$ in between each particle. \ Requiring finiteness of the
Hamiltonian at $x=\pm \infty $ and a consistent match of the solutions
across the delta-functions at the particle boundaries yields%
\begin{eqnarray}
L_{1}L_{2}L_{3} &=&{\frak M}_{12}{\frak M}_{21}L_{3}^{\ast }e^{\frac{\kappa 
}{4}s_{12}[(L_{1}+{\frak M}_{12})z_{13}-(L_{2}+{\frak M}_{21})z_{23}]} 
\nonumber \\
&&+{\frak M}_{23}{\frak M}_{32}L_{1}^{\ast }e^{\frac{\kappa }{4}%
s_{23}[(L_{2}+{\frak M}_{23})z_{21}-(L_{3}+{\frak M}_{32})z_{31}]}  \nonumber
\\
&&+{\frak M}_{31}{\frak M}_{13}L_{2}^{\ast }e^{\frac{\kappa }{4}%
s_{31}[(L_{3}+{\frak M}_{31})z_{32}-(L_{1}+{\frak M}_{13})z_{12}]}
\label{Htrans}
\end{eqnarray}%
where ${\frak M}_{ij}=M_{i}-\epsilon p_{i}s_{ij}$, $M_{i}=\sqrt{%
p_{i}^{2}+m_{i}^{2}}$, $L_{i}=H-M_{i}-\epsilon (\sum_{j}p_{j}s_{ji})$, $%
L_{i}^{\ast }=(1-\prod_{j<k\neq i}s_{ij}s_{ik})M_{i}+L_{i}$, $%
z_{ij}=(z_{i}-z_{j})$ and $s_{ij}=$sgn$(z_{ij})$. For simplicity we have set 
$\Lambda =0$. The discrete constant of integration $\epsilon =\pm 1$ flips
sign under time-reversal; it provides a measure of the flow of time of the
gravitational field relative to the particle momenta. \ 

Eq. (\ref{Htrans}) is an exact result and implicitly determines the
Hamiltonian $H$ as a function of the two independent coordinate degrees of
freedom $\left( \rho ,\lambda \right) $ and their conjugate momenta, which
can be written as 
\begin{eqnarray}
z_{1}-z_{2} &=&\sqrt{2}\rho \text{ \ \ \ \ \ \ \ \ \ \ }z_{1}+z_{2}-2z_{3}=%
\sqrt{6}\lambda  \label{hextrans1} \\
p_{1}-p_{2} &=&\sqrt{2}p_{\rho }\text{ \ \ \ \ \ \ }p_{1}+p_{2}-2p_{3}=\sqrt{%
6}p_{\lambda }  \label{hextrans2}
\end{eqnarray}%
choosing the center of momentum to vanish. The usual relations $\dot{z}_{a}=%
\frac{\partial H}{\partial p_{a}}$ and $\dot{p}_{a}=-\frac{\partial H}{%
\partial z_{a}}$ yield the equations of motion, and it is straightforward to
show that the Hamiltonian is time-independent when these equations hold. \ 

A post-Newtonian expansion \cite{ohtarobb} of (\ref{Htrans}) yields 
\begin{eqnarray}
H &=&3mc^{2}+\frac{p_{\rho }^{2}+p_{\lambda }^{2}}{2m}+\frac{\kappa m^{2}}{%
\sqrt{8}}\left[ \left| \rho \right| +\frac{\sqrt{3}}{2}\left( \left| \lambda
+\frac{\rho }{\sqrt{3}}\right| +\left| \lambda -\frac{\rho }{\sqrt{3}}%
\right| \right) \right] -\frac{(p_{\rho }^{2}+p_{\lambda }^{2})^{2}}{%
16m^{3}c^{2}}+\frac{\kappa c^{2}}{\sqrt{8}}|\rho |p_{\rho }^{2}  \nonumber \\
&&+\frac{\kappa c^{2}}{16\sqrt{2}}\left[ 3\sqrt{3}\left( \left| \lambda +%
\frac{\rho }{\sqrt{3}}\right| +\left| \lambda -\frac{\rho }{\sqrt{3}}\right|
\right) \left( p_{\lambda }^{2}+p_{\rho }^{2}\right) +6\left( \left| \lambda
+\frac{\rho }{\sqrt{3}}\right| -\left| \lambda -\frac{\rho }{\sqrt{3}}%
\right| \right) p_{\rho }p_{\lambda }\right]  \nonumber \\
&&+\frac{\kappa ^{2}m^{3}c^{6}}{16}\left[ \frac{\left| \rho \right| \sqrt{3}%
}{2}\left( \left| \lambda +\frac{\rho }{\sqrt{3}}\right| +\left| \lambda -%
\frac{\rho }{\sqrt{3}}\right| \right) +\frac{3}{4}\left| \lambda +\frac{\rho 
}{\sqrt{3}}\right| \left| \lambda -\frac{\rho }{\sqrt{3}}\right| -\frac{3}{4}%
\left( \lambda ^{2}+\rho ^{2}\right) \right]  \label{HpN}
\end{eqnarray}%
in the equal mass case. \ The first term is the total rest mass and the next
two terms are equivalent to the hexagonal-well Hamiltonian of a single
particle\cite{Butka}, which we refer to as the hex-particle. \ We see that
to leading order in $c^{-2}$ the hexagon potential is modified to have
parabolic edges, as well as a momentum-dependent steepening of its walls.
For unequal masses the hexagonal symmetry becomes distorted, with two
opposite corners moving inward, changing the relative arc-lengths of the
sides.{\bf \ }The full potential (obtained from (\ref{Htrans}) by setting \ $%
p_{\rho }=p_{\lambda }=0$) has the shape of a hexagonal carafe with convex
edges.

The system's interesting dynamics are due to the non-smoothness of the
potential along the lines $\rho =0$, $\rho -\sqrt{3}\lambda =0$, and $\rho +%
\sqrt{3}\lambda =0$, respectively corresponding to the crossings of \
particles 1 and 2, 2 and 3, or 1 and 3. We distinguish two distinct types of
motion \cite{Lmiller}: $A$-motion, corresponding to the same pair of
particles crossing twice in a row (the hex-particle crossing a single
sextant boundary twice in succession), and $B$-motion, in which one particle
crosses each of its compatriots in succession (the hex-particle crossing two
successive sextant boundaries). We can characterize a given motion by a
sequence of letters $A$ and $B$ (called a symbol sequence), with a finite
exponent $n$ denoting $n$-repeats and an overbar denoting an infinite
repeated sequence.

We numerically solve the equations of motion that follow from (\ref{Htrans})
in both the exact relativistic (R) system and its non-relativistic (N) limit
(ie $c\rightarrow \infty $). We impose absolute and relative error
tolerances of $10^{-8}$ for the numerical ODE solvers. This yields
numerically stable solutions for the values of $H\leq 6mc^{2}$ that we
study; beyond this the ODE solving algorithm is unstable. We test stability
by checking that the energy remains a constant of the motion to within a
value no larger than $10^{-6}$ throughout. \ 

For both N and R systems we find three principal classes of motion. One
class, with sequence $\overline{B}$,\ describes annulus-shaped orbits
encircling the origin in the $\rho -\lambda $\ plane. In the second
(pretzel) class the hex-particle essentially oscillates back and forth about
one of the three bisectors in a combination of $A$ and $B$ motions. Chaotic
orbits constitute the third class, where the hex-particle wanders between $A$%
- and $B$-motions in an apparently irregular fashion. Such orbits eventually
wander into all allowed areas of the $\rho -\lambda $ plane. \ 

\begin{figure}[tbp]
\begin{center}
\epsfig{file=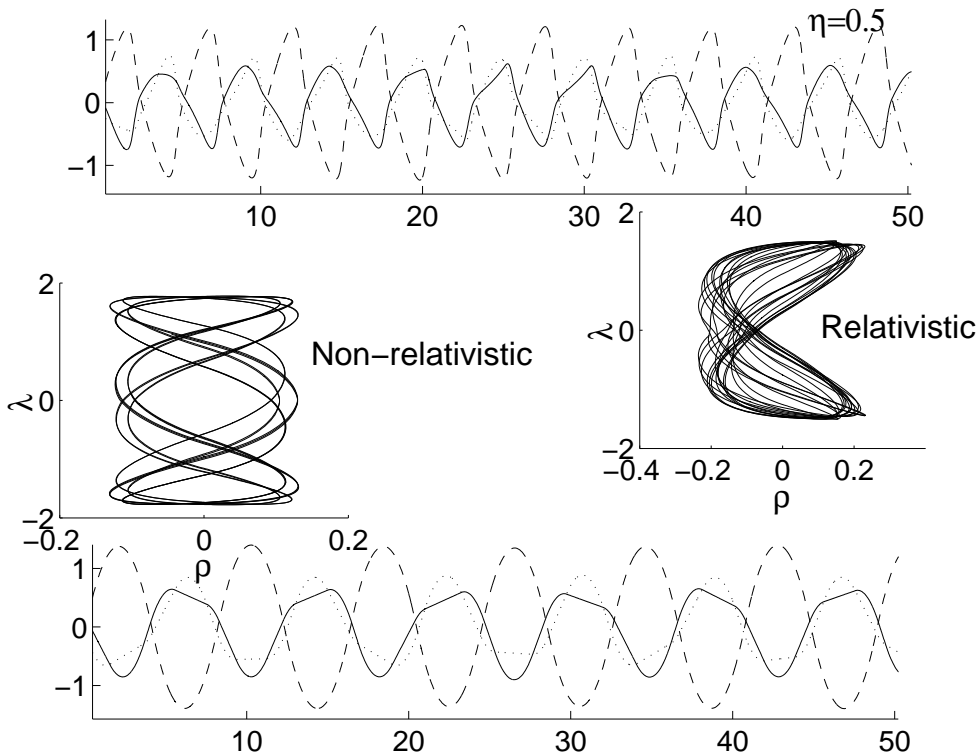,width=0.8\linewidth}
\end{center}
\caption{Regular pretzel orbits (run for 120 time steps) of the R and N
systems with the corresponding $3$-particle time evolution (truncated at 50
time steps). The collision sequences differ for these initial conditions and
correspond to $AB^{3}$ (R) and $A^{2}B^{3}$ (N). }
\label{3bdshortf1}
\end{figure}

Although the orbits of the N and R systems realize the same symbol
sequences, important qualitative differences exist between them. For a
meaningful comparison between each system we set the total energy $H=E_{T}$
and the initial values of $\left( \rho ,\lambda ,p_{\rho }\right) $ to be
the same, setting the initial value of $p_{\lambda }$ to satisfy (\ref%
{Htrans}) in each case. \ As the parameter \ $\eta =H/\left( 3mc^{2}\right)
-1$\ grows, we find that R trajectories have higher frequencies and extend
over a smaller region of the $\left( \rho ,\lambda \right) $ plane than
their N counterparts. \ Choosing initial conditions so that $A$-motion takes
place at $\lambda =0$\ we find that R trajectory patterns narrow with
increasing $\eta $\ in the small-$\lambda $\ region, whereas those for the N
system do not. Figure (\ref{3bdshortf1}) illustrates a typical example with $%
\eta =0.5$ for the pretzel case, in which the trajectories for the 3 bodies
and their corresponding hex-particle are shown. For each, two of the bodies
form a low-amplitude/high-frequency bound state that in turn oscillates at
higher amplitude and lower frequency with the third. The 3-body N
oscillations are parabolic in shape whereas there is shoulder-like
distortion in the R system due to the momentum-dependence of the potential
(a feature seen previously in the $2$-body case \cite{exact2body}) \
illustrating that under appropriate initial conditions two bodies can
tightly and stably bind together in both the N and R systems. The
hex-trajectories differ substantively, with the R pattern having 50\% higher
frequency, and extending over a slightly smaller range than its N
counterpart. The R-``boomerang'' also has an indentation on the left at $%
\lambda =0$\ that is absent for all corresponding N system boomerangs\ (not
shown). \ 

\begin{figure}[tbp]
\begin{center}
\epsfig{file=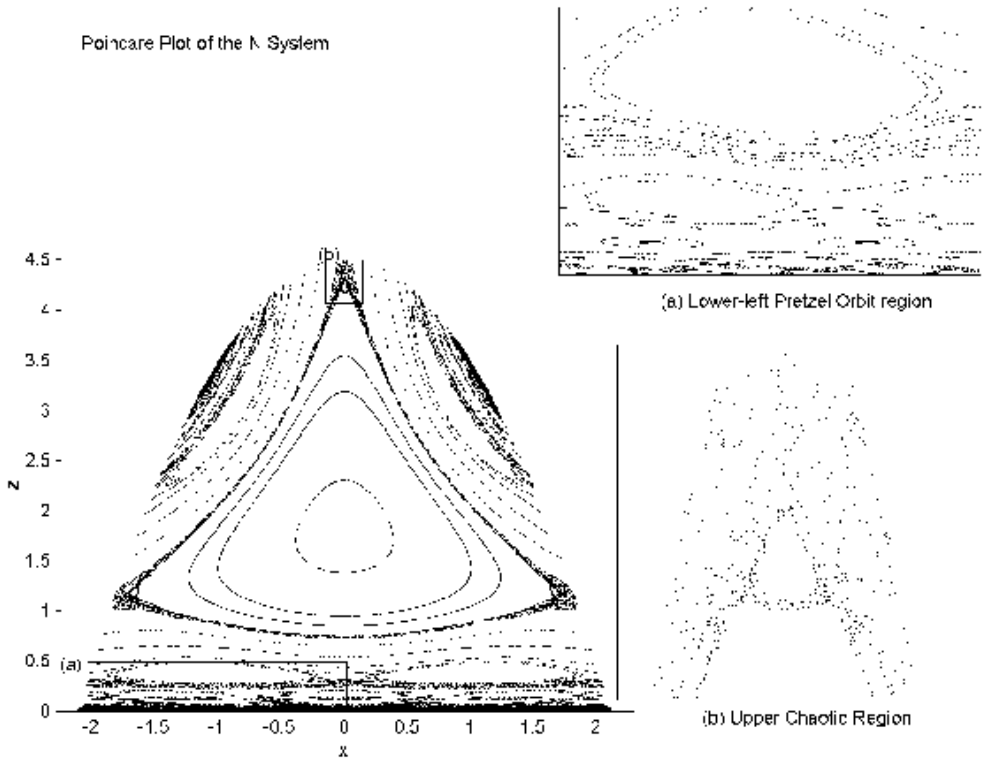,width=0.8\linewidth}
\end{center}
\caption{The Poincare plot of the N system. The squares denote the parts of
the plot magnified in the insets. }
\label{Poincare0b}
\end{figure}

\begin{figure}[tbp]
\begin{center}
\epsfig{file=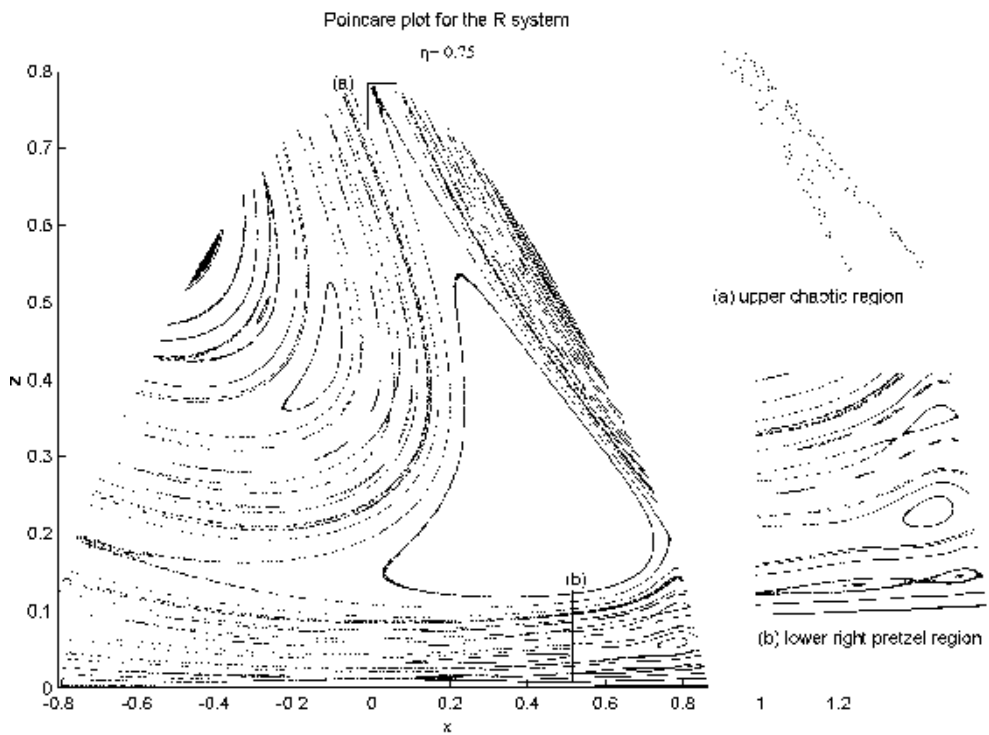,width=0.8\linewidth}
\end{center}
\caption{The Poincare plot of the R system at $\protect\eta =0.75$ and $%
\protect\epsilon =+1$. The upper right inset provides a close-up of the
chaotic region at the top of the diagram; it is now considerably narrower
than for lower values of $\protect\eta $. The lower-right inset is a
close-up of the structure in a pretzel region in the lower right of the
diagram. }
\label{Poincare175}
\end{figure}

We probe the global structure of the phase-space using Poincare plots, which
we construct by recording the radial $p_{r}$ and angular momentum $j$ each
time the hex-particle crosses a hexagonal bisector, plotting their values as
a series of dots in the $\left( p_{r},j^{2}\right) $ plane. Periodic
trajectories appear as a finite series of dots, quasi-periodic trajectories
as closed loops and chaotic ones as densely filled regions. Our numerical
results (see fig. \ref{Poincare0b}\ ) for the N system agree with previous
studies \cite{Butka,Lmiller,MillerReid}, which established that a region of
phase space exists in which the hex-particle circulates regularly around the
origin, bounded by a thin region of chaos outside of which there is a
regular structure of periodic and quasi-periodic orbits. Remarkably, we do
not observe a breakdown from regular to chaotic motion as $\eta $ increases,
despite the high degree of non-linearity in the R system. The lower regions
of the Poincare map clearly display the same pattern of series of circles as
occurs in the non-relativistic case. However the Poincare plot for the R
system is no longer symmetric with respect to $p_{r}=0$, but instead is
`squashed' towards the right-hand side (compare fig. \ref{Poincare0b}\ to
fig.\ref{Poincare175}). This deformation is reminiscent of the situation for
two particles, in which the gravitational coupling to the kinetic-energy of
the particles causes a distortion of the trajectory from an otherwise
symmetric pattern \cite{exact2body}. Since (\ref{Htrans}) is invariant only
under the discrete symmetry $\left( p_{i},\epsilon \right) \rightarrow
\left( -p_{i},-\epsilon \right) $ (and not under $p_{i}\rightarrow -p_{i}$),
we \ obtain for the choice $\epsilon =\pm 1$ a distortion towards the lower
right/left-hand side of the plot relative to its N counterpart.

No sizeable connected areas of chaos are present in either the R or N
systems, indicating that most trajectories are effectively restricted to
move on two-dimensional surfaces in phase space. We find for all $\eta $
within the range investigated that chaotic orbits exist separating the
annulus and pretzel regions. We therefore conjecture that the R system is
not integrable, since its N counterpart is known to be non-integrable \cite%
{Lmiller}. Nonetheless, clearly some underlying feature enforces
considerable structure on the phase space, preventing KAM-breakdown to
global chaos typical of such Hamiltonian systems \cite{KAM}.

In general pretzel-type orbits display a remarkable richness of dynamics. As
the initial angular momentum of the trajectory in question decreases, the
number of successive $A$\ collisions increases before the hex-particle
sweeps around the origin in a $B^{3}$-sequence, corresponding to a
180-degree swing of the hex-particle around the origin . \ For example $%
\overline{AB^{3}}$ (the simplest sequence after $\overline{B}$) corresponds
to a boomerang-type orbit and appears as two circles on the Poincare
section. The next simplest sequence is $\overline{A^{2}B^{3}}$, which
corresponds to a bow-tie like orbit, and generates three slightly smaller
circles on the Poincare section. For each of these patterns a family of
orbits exists corresponding to different widths of the `bands' of phase
space that the trajectory covers, and correspondingly different radii of
circles in the Poincare section. Between these regions, the orbits'
sequences are mixtures of $AB^{3}$\ and $A^{2}B^{3}$. This reasoning can be
extended to more general $A^{n}B^{3m}$-motions. We conjecture that the only
allowed non-chaotic orbits -- relativistic and non-relativistic -- are of
the form $\prod_{n,m,j}\left( A^{n}B^{3m}\right) ^{l_{j}}$\ with $n,m$\
finite, corresponding to increasingly complex weaving patterns.

If the set of integers $l_{j}$\ is finite, then the sequence is regular,
leaving bands of phase space untravelled, and appearing as a series of
closed crescents or ellipsoids on the Poincare section. If, however, the
sequence of integers $l_{j}$\ never repeats itself, then the trajectory will
fill the available phase space densely, appearing as a wavy line on the
surface of section. We conjecture that there is a $1-1$ correspondence
between rational numbers and periodic orbits in this region of phase space,
both non-relativistically and relativistically. This would give the lower
section of the Poincare plot a fractal structure as the patterns of circles,
ellipses and lines is repeated on arbitrarily small scales as the
hex-particle's angular momentum approaches zero.

The 3-body relativistic system provides a new theoretical laboratory for
studying the interplay between relativity and chaos. While the dynamical
structure of the relativisitc system bears some resemblance to well-known
features of its non-relativistic counterpart (such as the self-similar
structure of the pretzel class and the lack of energy dependence of the
orbit topology), significant differences arise in the shape and frequency of
the orbits.\ A full study of the large-$\eta $ regime should provide us with
interesting new information to this end.

\bigskip

This work was supported by the Natural Sciences and Engineering Research
Council of Canada.

\end{document}